\documentclass[aps,prd,amsmath,amssymb,nofootinbib,longbibliography]{revtex4-2}

\usepackage{graphicx}
\usepackage{xcolor}
\usepackage{dcolumn}
\usepackage{bm}
\usepackage{amsthm,physics,hyperref,mathtools,dsfont}
\usepackage{yhmath}
\usepackage[capitalise]{cleveref}
\usepackage{bbm}

\begin{document}

\title{Bias in Local Spin Measurements from Deformed Symmetries}

\author{Michele Arzano}
\email{michele.arzano@unina.it}

\affiliation{Dipartimento di Fisica ``E. Pancini", Universit\`a di Napoli Federico II, I-80125 Napoli, Italy}
\affiliation{INFN, Sezione di Napoli, Complesso Universitario di Monte S. Angelo,
Via Cintia Edificio 6, 80126 Napoli, Italy}

\author{Goffredo Chirco}
\email{goffredo.chirco@unina.it}

\affiliation{Dipartimento di Fisica ``E. Pancini", Universit\`a di Napoli Federico II, I-80125 Napoli, Italy}
\affiliation{INFN, Sezione di Napoli, Complesso Universitario di Monte S. Angelo,
Via Cintia Edificio 6, 80126 Napoli, Italy}

\author{Jerzy Kowalski-Glikman}
\email{jerzy.kowalski-glikman@ncbj.gov.pl}

\affiliation{National Centre for Nuclear Research, Pasteura 7, 02-093 Warsaw, Poland}
\affiliation{Faculty of Physics and Astronomy, University of Wroclaw, Pl. Maksa Borna 9, 50-204
Wroclaw, Poland}

\date{\today}

\begin{abstract}

We study local spin measurements on bipartite singlet states when rotational symmetry is described by a quantum group rather than an ordinary Lie group. Although the fundamental spin-\(\frac12\) representation has the usual one-particle action on states, the non-trivial coproduct selects a deformed analogue of the Bell singlet state. We show that conventional tensor-factor measurements on this invariant singlet lead to deformation-dependent one-site outcome statistics. We then compare this standard-local prescription with a
braided-local one, obtained by dressing local observables with the \(R\)-matrix and using the corresponding braided inner product. The braided
observables are covariant under the Hopf adjoint action, but define a distinct measurement structure and lead to the reciprocal bias on the deformed singlet state. 
\end{abstract}

\maketitle

Rotational symmetry and the associated conservation of angular momentum constitute one of the most fundamental pillars of physics. But what if this symmetry is subtly modified at a fundamental level? Any such modification
must be exceedingly small, as no deviations from exact rotational symmetry have been observed in quantum mechanical or quantum field--theoretical systems to date.

The situation may, however, be different in the context of quantum gravity. Indeed, over the past decades, increasing evidence  suggests that, upon quantization, the local symmetry group of gravity may acquire a richer algebraic structure. This structure goes beyond the realm of ordinary Lie groups and Lie algebras and is naturally encoded in the framework of Hopf algebras, or quantum groups \cite{ChariPressley1994,Majid:1996kd}. In lower dimensions this phenomenon is particularly well understood: in \(2+1\) dimensions the Einstein--Hilbert action admits a Chern--Simons formulation, and the
quantization of the theory naturally leads to quantum-group structures \cite{Bais:1998yn,Bais:2002ye,Meusburger:2003hc,Meusburger:2003ta,Cianfrani:2016ogm}. In \(3+1\) dimensions the situation is more subtle, because of the presence of local gravitational degrees of freedom, but several independent lines of evidence suggest that Hopf-algebraic
symmetries may arise once quantum-gravity effects and a cosmological constant are taken into account
\cite{Amelino-Camelia:2003ezw,Kowalski-Glikman:2008fix,Arzano:2021scz} (see also \cite{Addazi:2021xuf} for a comprehensive discussion of the phenomenological and observational consequences of such modifications).

For the purposes of the present work, the important point is not the precise microscopic origin of the deformation, but the structural fact that Hopf-algebraic symmetries modify the way symmetry
acts on composite systems. In an ordinary Lie algebra the action on composite systems satisfies the Leibniz rule, and it is compatible with the usual tensor-product notion of locality. In a quantum group setting, by contrast, the action on composite systems is generally non-leibnizian. As a result, the action of symmetry on composite systems and the embedding of local observables into a bipartite Hilbert space are no longer related in the standard way.

We therefore study the simplest setting in which this effect can be seen explicitly: a pair of spin-\(\frac12\) systems with rotational symmetry
described by the Hopf algebra \(U_q(\mathfrak{su}(2))\) \cite{biedenharn1995quantum}.\footnote{See \cite{Amelino-Camelia:2022dsj} for a proposal to use the quantum group \(SU_q(2)\), the ``group'' counterpart of \(U_q(\mathfrak{su}(2))\), to describe quantum angular reference frames.} We take the deformation parameter to be real and positive, \(q\in\mathbb R_{>0}\),
and use \(U_q(\mathfrak{su}(2))\) as a concrete model of deformed rotational symmetry\footnote{In positive-\(\Lambda\) approaches, quantum-group structures are often
associated with compact or root-of-unity regimes and with finite spin or angular cutoffs; see for example \cite{Bianchi:2011uq}. This provides an important motivation for quantum-group deformations in quantum gravity, but it is not the regime used in the present calculation, where we work with real \(q>0\). } The aim is not to derive \(q\) from a specific quantum-gravity model, but to isolate the operational consequences of a
non-trivial Hopf algebra structure on bipartite systems. 

The Hopf algebra
\(U_q(\mathfrak{su}(2))\) is generated by \(J_z\), \(J_+\) and \(J_-\),
subject to the commutation relations
\begin{equation}
[J_z,J_\pm] = \pm J_\pm ,
\end{equation}
\begin{equation}
[J_+,J_-] =
\frac{q^{2J_z}-q^{-2J_z}}{q-q^{-1}},
\end{equation}
which in the limit \(q\to1\) reproduce the standard
\(\mathfrak{su}(2)\) Lie algebra.

An interesting property of this algebra \cite{KlimykSchmudgen1997,Arzano:2025zsp,Arzano:2026gng} is that in the fundamental spin-\(\frac12\) representation the one-particle action of the generators is undeformed. Indeed, for the \(J_z\)-eigenstates \(\ket{\uparrow}\) and \(\ket{\downarrow}\),
\begin{equation}
J_z|\uparrow\rangle=\tfrac12|\uparrow\rangle,
\qquad
J_z|\downarrow\rangle=-\tfrac12|\downarrow\rangle,
\qquad
J_+|\downarrow\rangle=|\uparrow\rangle,
\qquad
J_-|\uparrow\rangle=|\downarrow\rangle,
\end{equation}
we find
\begin{align}
    [J_+,J_-]\ket{\uparrow}
    &=
    \frac{q^{2J_z}-q^{-2J_z}}{q-q^{-1}}\ket{\uparrow}
    =
    \ket{\uparrow},
    \\
    [J_+,J_-]\ket{\downarrow}
    &=
    \frac{q^{2J_z}-q^{-2J_z}}{q-q^{-1}}\ket{\downarrow}
    =
    -\ket{\downarrow},
\end{align}
with all other actions vanishing. In this representation the algebra action coincides with that of ordinary \(\mathfrak{su}(2)\). Does this mean that, for a spin-\(\frac12\) system, the deformation has no observable
consequences? Not at all. The algebra is only one of the structures that characterize Hopf-algebra deformations of Lie algebras. Another structure of central importance in this paper is the {\it coproduct}.

The coproduct is a map which determines how the algebra elements act on tensor-product representations or, in physical terms, how the symmetry generators/quantum observables act on composite systems. In the undeformed case, the standard (``primitive") coproduct dictates a Leibniz action of symmetry generators on tensor product representations and is compatible with the usual tensor-factor notion of locality: local observables such as \(A\otimes\mathbbm 1\) transform as expected under the total symmetry action. For a Hopf algebra with a deformed coproduct this compatibility is no longer automatic. The total symmetry action, the invariant singlet state, and the embedding of local observables into the two-particle Hilbert space must be considered together.

This observation places our discussion in contact with familiar questions in quantum foundations and quantum information. The EPR argument and Bell's analysis already show that the physical content of a bipartite quantum state is inseparable from the specification of local measurements \cite{Einstein:1935rr,Bell:1964kc}.
In modern quantum information theory this point is sharpened further: entanglement, separability and LOCC protocols are defined relative to a tensor-product structure and to a distinguished class of local operations and measurements \cite{Horodecki:2009zz}. Moreover, the tensor-product structure itself can be regarded as induced by the accessible observables
\cite{Zanardi:2003zz}, and entanglement can be generalized as a property relative to a preferred algebra of observables rather than to a fixed
subsystem decomposition \cite{Barnum:2004zz}. Similarly, restrictions arising from reference frames and superselection rules modify the set of physically available operations hence the operational content of
quantum-information protocols \cite{BartlettRudolphSpekkens2007,KitaevMayersPreskill2004}.

We therefore ask a deliberately simple question: what becomes of the standard two-spin singlet measurement
experiment when rotational symmetry is described by
\(U_q(\mathfrak{su}(2))\)? Since the fundamental spin-\(\frac12\) representation is undeformed, one might expect
nothing observable to happen at the level of two qubits. We show that this expectation is misleading. The deformation is hidden in the single qubit
action of the generators, but it reappears in composite systems, in the Hopf adjoint action on observables, and in the distinction between ordinary tensor-factor locality and braided local measurements.

To make these statements concrete, we now focus on the Hopf-algebraic structure of \(U_q(\mathfrak{su}(2))\). The key ingredient, beyond the single-particle action displayed above, is the coproduct which determines the total action of the deformed symmetry on a pair of spin-\(\frac12\). With our conventions, the coproduct of the generators is
\begin{align}
\Delta(J_z) &= J_z\otimes \mathbbm{1} + \mathbbm{1}\otimes J_z ,\nonumber \\ 
\Delta(J_\pm) &= J_\pm\otimes q^{J_z} + q^{-J_z}\otimes J_\pm ,\label{copropm}
\end{align}
so, for example acting on the two-particle state $\ket{\uparrow\downarrow}= \ket{\uparrow}\otimes \ket{\downarrow}$ they give
\begin{align}
\Delta(J_z) \ket{\uparrow\downarrow} &=\Delta(J_z) \ket{\downarrow\uparrow}=0\,,\nonumber\\
\Delta(J_+)\ket{\uparrow\downarrow} &= q^{-1/2}\ket{\uparrow\uparrow}\,,
\quad
\Delta(J_+)\ket{\downarrow\uparrow} = q^{1/2}\ket{\uparrow\uparrow}\nonumber\\
\Delta(J_-)\ket{\uparrow\downarrow} &= q^{-1/2}\ket{\downarrow\downarrow}\,,
\quad
\Delta(J_-)\ket{\downarrow\uparrow} = q^{1/2}\ket{\downarrow\downarrow}.
\end{align}
Let us now consider the singlet two-particle state. Since a singlet is a state with vanishing total angular momentum it must be annihilated by all the three operators $\Delta J_z$, $\Delta J_\pm$.  We therefore seek a linear combination
\begin{equation}
|\psi\rangle = a\,|\uparrow\downarrow\rangle + b\,|\downarrow\uparrow\rangle
\end{equation}
which is annihilated by $\Delta J_\pm$. Using \eqref{copropm} one finds that the condition $\Delta(J_+)|\psi\rangle=0$ implies
\begin{equation}
a q^{-1/2} + b q^{1/2} = 0,
\qquad\text{or equivalently}\qquad
b=-q^{-1}a.
\end{equation}
The same relation follows from imposing $\Delta(J_-)|\psi\rangle=0$. We thus find that, upon appropriate normalization, the state annihilated by the action of $\Delta J_z$, $\Delta J_\pm$ has the form
\begin{equation}\label{qsingl}
\ket{\psi_q} = \frac{1}{\sqrt{1+q^{-2}}}\left(\ket{\uparrow\downarrow} - q^{-1}\ket{\downarrow\uparrow}\right).
\end{equation}
The central quadratic Casimir of $U_q(\mathfrak{su}(2))$ is given by
\begin{equation}
\label{eq:Uqsu2CasimirAlternative}
C_q
=
J_-J_+
+
[J_z]_q[J_z+1]_q ,
\end{equation}
where 
\begin{equation}
[x]_q \equiv \frac{q^x-q^{-x}}{q-q^{-1}} .
\end{equation}
The eigenvalue of $C_q$ on the spin-$j$ irreducible representation is
\begin{equation}
C_q\big|_j = [j]_q[j+1]_q .
\end{equation}
The Casimir on tensor product states is obtained by using the coproducts 
\begin{equation}
\label{eq:twoQubitCasimirSymmetric}
J_{\mathrm{tot}}^2=C_q^{(12)}
=
\frac12
\left(
\Delta J_+\Delta J_-
+
\Delta J_-\Delta J_+
\right)
+
\frac{[2]_q}{2}
[\Delta J_z]_q^2 ,
\end{equation}
with $[2]_q=q+q^{-1}$, and we thus have
\begin{equation}
J_{\mathrm{tot}}^2\,\ket{\psi_q}  = 0.
\end{equation}
Thus the deformed Bell state $\ket{\psi_q} $ carries total spin $j=0$ and realizes the trivial representation of $U_q(\mathfrak{su}(2))$. 

This result illustrates a general feature of quantum group deformations: while the local action on a single spin-$\tfrac12$ system is undeformed, the coproduct modifies the structure of the action on tensor product representations (see \cite{Borowiec:2009vb,Arzano:2021scz} for a deformation of the Poincaré algebra with similar features i.e. undeformed commutators and non-trivial coproducts). In particular, the singlet state persists as an invariant vector, but its explicit form acquires a $q$-dependent deformation, whereas the naive antisymmetric combination is invariant only in the undeformed limit $q\rightarrow1$.

For an ordinary \(\mathfrak{su}(2)\) Lie algebra the operator
\begin{equation}\label{loc1m}
 J_a^{(A)} \equiv J_a\otimes \mathbbm{1}
\end{equation}
represents a local spin measurement performed by Alice on the first subsystem. The coproduct of the undeformed Lie algebra is primitive,
\begin{equation}\label{eq:primitivecoproduct}
\Delta(X)
=
X\otimes \mathbbm{1}
+
\mathbbm{1}\otimes X,
\qquad
X\in\mathfrak{su}(2),
\end{equation}
so that the two-spin representation is given by
\begin{equation}\label{rho2cpr}
\rho^{(2)}(X)
:=
(\rho\otimes\rho)\Delta(X)
=
\rho(X)\otimes\mathbbm{1}
+
\mathbbm{1}\otimes\rho(X),
\end{equation}
where $\rho(X)$ is the one qubit representation of $X\in\mathfrak{su}(2)$. At the finite group level, for \(g\in \mathrm{SU}(2)\), the corresponding
two-particle representation is
\begin{equation}
\rho^{(2)}(g)=\rho(g)\otimes\rho(g),
\end{equation}
and the local observable \eqref{loc1m} is covariant:
\begin{equation}\label{eq:classicalfinitecovariance}
\rho^{(2)}(g)\,
(J_a\otimes\mathbbm{1})\,
\rho^{(2)}(g)^{-1}
=
\bigl(\rho(g)J_a\rho(g)^{-1}\bigr)
\otimes\mathbbm{1}.
\end{equation}
Equivalently, infinitesimally one has
\begin{equation}\label{eq:classicalinfinitesimalcovariance}
\left[
\rho^{(2)}(X),
J_a\otimes\mathbbm{1}
\right]
=
\left[
\rho(X),J_a
\right]\otimes\mathbbm{1},
\qquad
X\in\mathfrak{su}(2).
\end{equation}
In particular, taking \(X=J_b\), this gives
\begin{equation}\label{adjord}
\left[
\rho^{(2)}(J_b),
J_a\otimes\mathbbm{1}
\right]
=
[J_b,J_a]\otimes\mathbbm{1},
\end{equation}
where we suppress the representation symbol on the single-spin generators. Thus, in the undeformed case, ``transform then localize'' and ``localize then transform'' are equivalent operations.\\

In the case of a deformed symmetry described by a Hopf algebra, this structure changes. Although the operators \(J_a\otimes\mathbbm 1\) and \(\mathbbm 1\otimes J_a\) remain well-defined Hermitian operators in the
standard tensor-product Hilbert space, the deformed coproduct does not transform Alice's and Bob's tensor-factor observables symmetrically. With the chosen ordering of the coproduct, Alice's observables transform among themselves under the Hopf adjoint action, whereas Bob's observables acquire a dressing on Alice's factor. Let us show this explicitly.

We first recall that, besides the coproduct, a Hopf algebra is equipped with an antipode map \(S\), which plays the role of an algebraic inverse. Writing
\[
K=q^{J_z},
\]
we have
\begin{equation}
S(J_z)=-J_z,
\qquad
S(K)=K^{-1},
\end{equation}
and
\begin{equation}
S(J_+)=-qJ_+,
\qquad
S(J_-)=-q^{-1}J_- .
\end{equation}
Notice that the element \(K\) is {\it group-like} i.e. its coproduct is given by 
\begin{equation}
\Delta K=K\otimes K.
\end{equation}

Together with the coproduct, the antipode allows one to define the Hopf analogue of the adjoint action. Let \(\rho\) be the one-qubit Hilbert-space representation of \(U_q(\mathfrak{su}(2))\). If \(O\in\mathrm{End}(\mathbb C^2)\) is a one-qubit operator, the Hopf adjoint
action of \(h\in U_q(\mathfrak{su}(2))\) on \(O\) is
\begin{equation}
\operatorname{ad}^{(1)}_h(O)
=
\rho(h_{(1)})\,O\,\rho(S h_{(2)}),
\label{eq:HopfAdjointOneQubit}
\end{equation}
where we used Sweedler notation
\begin{equation}
\Delta h=h_{(1)}\otimes h_{(2)}
\end{equation}
with summation understood. For \(J_z\), whose coproduct is primitive, this
reduces to the ordinary commutator:
\begin{equation}
\operatorname{ad}^{(1)}_{J_z}(O)
=
[\rho(J_z),O].
\end{equation}
For \(J_\pm\), instead, the non-primitive coproduct gives
\begin{equation}
\operatorname{ad}^{(1)}_{J_+}(O)
=
\rho(J_+)O\rho(K^{-1})
-
q\,\rho(K^{-1})O\rho(J_+),
\end{equation}
and
\begin{equation}
\operatorname{ad}^{(1)}_{J_-}(O)
=
\rho(J_-)O\rho(K^{-1})
-
q^{-1}\rho(K^{-1})O\rho(J_-).
\end{equation}
Thus, even though the action of the generators on spin-\(\frac12\) states is undeformed, the Hopf adjoint action on one-qubit observables is already \(q\)-dependent (see \cite{Arzano:2026gng} for a discussion of the physical interpretation of this fact in the context of the dynamics of qubits). For example one finds 
\begin{equation}
\operatorname{ad}^{(1)}_{J_+}(J_z)=-q^{1/2}\,J_+,
\qquad
\operatorname{ad}^{(1)}_{J_-}(J_z)=q^{-1/2}J_-\,.
\end{equation}

On a two-qubit system the relevant representation is instead induced by the coproduct,
\begin{equation}
\rho^{(2)}(h):=(\rho\otimes\rho)\Delta(h),
\qquad
h\in U_q(\mathfrak{su}(2)).
\end{equation}
If \(T\in\mathrm{End}(\mathbb C^2\otimes\mathbb C^2)\) is a two-qubit
operator, the corresponding Hopf adjoint action is
\begin{equation}
\operatorname{ad}^{(2)}_h(T)
=
\rho^{(2)}(h_{(1)})\,T\,\rho^{(2)}(S h_{(2)}).
\label{eq:HopfAdjointTwoQubit}
\end{equation}
Again, for \(J_z\) this reduces to the ordinary commutator,
\begin{equation}
\operatorname{ad}^{(2)}_{J_z}(T)
=
[\rho^{(2)}(J_z),T],
\end{equation}
whereas for \(J_\pm\) one obtains
\begin{equation}
\operatorname{ad}^{(2)}_{J_+}(T)
=
\rho^{(2)}(J_+)T\rho^{(2)}(K^{-1})
-
q\,\rho^{(2)}(K^{-1})T\rho^{(2)}(J_+),
\end{equation}
and
\begin{equation}
\operatorname{ad}^{(2)}_{J_-}(T)
=
\rho^{(2)}(J_-)T\rho^{(2)}(K^{-1})
-
q^{-1}\rho^{(2)}(K^{-1})T\rho^{(2)}(J_-).
\end{equation}
Here
\begin{equation}
\rho^{(2)}(K^{-1})
=
\rho(K^{-1})\otimes\rho(K^{-1}),
\end{equation}
because \(K\) is group-like.\\

Let us now consider the local Alice and Bob observables 
\begin{equation}
J_a^{(A)}=J_a\otimes 1,
\qquad
J_a^{(B)}=1\otimes J_a.
\end{equation}
We suppress the representation symbol on the one-qubit generators and we set
\begin{equation}
r=q^{1/2},
\qquad
s=r+r^{-1},
\qquad
d=r^{-1}-r .
\end{equation}
For the left-local observable \(J_z^{(A)}\), one finds
\begin{equation}
\mathrm{ad}^{(2)}_{J_+}(J_z^{(A)})
= -r\,J_+\otimes \mathbbm{1} = -r\,J_+^{(A)},
\end{equation}
and
\begin{equation}
\mathrm{ad}^{(2)}_{J_-}(J_z^{(A)})
= r^{-1}J_- \otimes \mathbbm{1} = 
r^{-1}J_-^{(A)}.
\end{equation}
Therefore, defining
\begin{equation}
J_x=\frac{J_++J_-}{2},
\qquad
J_y=\frac{J_+-J_-}{2i},
\end{equation}
and using linearity of the adjoint action in the acting Hopf-algebra element,
\begin{equation}
\mathrm{ad}^{(2)}_{J_x}(J_z^{(A)})
=
\frac{d}{2}J_x^{(A)}
-
\frac{is}{2}J_y^{(A)},
\end{equation}
while
\begin{equation}
\mathrm{ad}^{(2)}_{J_y}(J_z^{(A)})
=
\frac{is}{2}J_x^{(A)}
+
\frac{d}{2}J_y^{(A)}.
\end{equation}
More generally, the left-local algebra generated by
\(\{1,J_x^{(A)},J_y^{(A)},J_z^{(A)}\}\) is stable under this left Hopf adjoint action.

For the right-local observable \(J_z^{(B)}\), one instead obtains
\begin{equation}
\mathrm{ad}^{(2)}_{J_+}(J_z^{(B)})
=
-r\,K^{-2}\otimes J_+,
\end{equation}
and
\begin{equation}
\mathrm{ad}^{(2)}_{J_-}(J_z^{(B)})
=
r^{-1}K^{-2}\otimes J_-.
\end{equation}
Hence we have
\begin{equation}
\mathrm{ad}^{(2)}_{J_x}(J_z^{(B)})
=
K^{-2}\otimes
\left(
\frac{d}{2}J_x
-
\frac{is}{2}J_y
\right),
\end{equation}
and
\begin{equation}
\mathrm{ad}^{(2)}_{J_y}(J_z^{(B)})
=
K^{-2}\otimes
\left(
\frac{is}{2}J_x
+
\frac{d}{2}J_y
\right).
\end{equation}
Thus, with the chosen coproduct ordering, the left-local algebra is stable under the left Hopf adjoint action, whereas the strict right-local algebra is not: it acquires a \(K^{-2}\) dressing on the first tensor factor. This ordering asymmetry shows that the naive tensor-factor notion of locality is not a symmetric Hopf-covariant structure once the coproduct is non-cocommutative.\\

A natural way to formulate symmetry-compatible local observables is instead to use a {\it braided} embedding of a single-site observable $A$ obtained by a $R$-matrix dressing \cite{Fiore:2000qj},
\begin{equation}\label{dressedA}
\widetilde{A}
=
(\rho \otimes \rho)(R_{21})\,
(A \otimes \mathbbm{1})\,
(\rho \otimes \rho)(R_{21}^{-1}),
\end{equation}
where $R_{21}=PRP$ with $P$ the tensor flip operator
\begin{equation}
P(v\otimes w)=w\otimes v,
\end{equation}
which in the basis $\mathcal B=\bigl\{
|\uparrow\uparrow\rangle,\,
|\uparrow\downarrow\rangle,\,
|\downarrow\uparrow\rangle,\,
|\downarrow\downarrow\rangle
\bigr\}$ reads
\begin{equation}
P=
\begin{pmatrix}
1&0&0&0\\
0&0&1&0\\
0&1&0&0\\
0&0&0&1
\end{pmatrix},
\end{equation}
and $R$ is the R-matrix satisfying the quasitriangularity relation
\begin{equation}\label{quasitid}
\Delta^{\mathrm{op}}(h)
=
R \Delta(h) R^{-1},
\end{equation}
with $\Delta^{\mathrm{op}} := P \circ \Delta$. In the ordered basis $\mathcal B$ we have explicitly \cite{biedenharn1995quantum,Tjin:1991me}

\begin{equation}\label{eq:Rfund}
\mathcal{R}\equiv(\rho \otimes \rho) R
=
\begin{pmatrix}
q & 0 & 0 & 0\\
0 & 1 & q-q^{-1} & 0\\
0 & 0 & 1 & 0\\
0 & 0 & 0 & q
\end{pmatrix}.
\end{equation}
Therefore
\begin{equation}\label{eq:R21}
\mathcal{R}_{21}\equiv (\rho \otimes \rho) R_{21}
=
\begin{pmatrix}
q & 0 & 0 & 0\\
0 & 1 & 0 & 0\\
0 & q-q^{-1} & 1 & 0\\
0 & 0 & 0 & q
\end{pmatrix}.
\end{equation}
Equivalently, one could multiply both \(R\) and \(R_{21}\) by an overall nonzero scalar without affecting the dressed operator \(\widetilde A\). Let us now go back to the local operator on the first factor,
\begin{equation}
J_z^{(A)}=J_z\otimes\mathbbm 1
=
\mathrm{diag}\!\left(
\tfrac12,\,
\tfrac12,\,
-\tfrac12,\,
-\tfrac12
\right)
\qquad \text{on }\mathcal B.
\end{equation}
The \(R\)-dressed embedding is
\begin{equation}\label{eq:dresseddef}
\widetilde J_z^{(A)}
=
\mathcal{R}_{21}\,
(J_z\otimes\mathbbm 1)\,
\mathcal{R}_{21}^{-1}.
\end{equation}
Since \(\mathcal{R}_{21}\) in \eqref{eq:R21} is lower triangular with a single
off-diagonal entry in the \((3,2)\) position, its inverse is
\begin{equation}
\mathcal{R}_{21}^{-1}
=
\begin{pmatrix}
q^{-1} & 0 & 0 & 0\\
0 & 1 & 0 & 0\\
0 & -(q-q^{-1}) & 1 & 0\\
0 & 0 & 0 & q^{-1}
\end{pmatrix}.
\end{equation}
A direct matrix multiplication in \eqref{eq:dresseddef} gives
\begin{equation}\label{eq:dressedmatrixA}
\widetilde J_z^{(A)}
=
\begin{pmatrix}
\frac12 & 0 & 0 & 0\\
0 & \frac12 & 0 & 0\\
0 & q-q^{-1} & -\frac12 & 0\\
0 & 0 & 0 & -\frac12
\end{pmatrix}.
\end{equation}
or, in more compact form
\begin{equation}\label{eq:dressedopform}
\widetilde J_z^{(A)}
=
J_z\otimes\mathbbm 1
+
\bigl(q-q^{-1}\bigr)\,
\sigma_-\otimes\sigma_+.
\end{equation}
Notice how this dressed operator {\it is not} hermitian with respect to the standard inner product on $\mathbb{C}^2 \otimes \mathbb{C}^2$, we will get back to this point soon.

In the undeformed limit $R \to \mathbbm{1} \otimes \mathbbm{1}$, the dressing disappears and a generic dressed generator $\widetilde{A}$ reduces to the usual local observable. Using the quasitriangular identity \eqref{quasitid}, one can verify that a generic dressed operator is covariant under the adjoint action of $U_q(\mathfrak{su}(2))$. For $h\in U_q(\mathfrak{su}(2))$ the two-qubit representation is given by \eqref{rho2cpr}
\begin{equation}
\rho^{(2)}(h):=(\rho\otimes\rho)\Delta(h),
\end{equation}
while we denote the opposite-coproduct representation by
\begin{equation}
\rho^{(2)}_{\mathrm{op}}(h):=(\rho\otimes\rho)\Delta^{\mathrm{op}}(h).
\end{equation}
Starting from the quasitriangularity relation \eqref{quasitid} or equivalently
\begin{equation}
R\Delta(h)=\Delta^{\mathrm{op}}(h)R,
\end{equation}
and using \(R_{21}=PRP\), one obtains the relation
\begin{equation}
R_{21}\Delta^{\mathrm{op}}(h)=\Delta(h)R_{21},
\qquad
h\in U_q(\mathfrak{su}(2)),
\end{equation}
which follows by conjugating the previous equation with the tensor flip \(P\), since \(P\Delta(h)P=\Delta^{\mathrm{op}}(h)\). In the two-qubit
representation this becomes
\begin{equation}
\mathcal R_{21}\,\rho^{(2)}_{\mathrm{op}}(h)
=
\rho^{(2)}(h)\,\mathcal R_{21}.
\end{equation}
For example, for \(J_+\) one has explicitly
\begin{equation}
\rho^{(2)}(J_+)
=
\begin{pmatrix}
0&q^{-1/2}&q^{1/2}&0\\
0&0&0&q^{-1/2}\\
0&0&0&q^{1/2}\\
0&0&0&0
\end{pmatrix},
\end{equation}
whereas
\begin{equation}
\rho^{(2)}_{\mathrm{op}}(J_+)
=
\begin{pmatrix}
0&q^{1/2}&q^{-1/2}&0\\
0&0&0&q^{1/2}\\
0&0&0&q^{-1/2}\\
0&0&0&0
\end{pmatrix}.
\end{equation}

On the \(R\)-dressed embedding of a one-qubit observable
$\widetilde A
=
\mathcal R_{21}\,
(A\otimes\mathbbm 1)\,
\mathcal R_{21}^{-1}$, using the quasitriangular relation, we have
\begin{equation}
\begin{split}
\rho^{(2)}(h)\,\widetilde A
&=
\rho^{(2)}(h)\,
\mathcal R_{21}(A\otimes\mathbbm 1)\mathcal R_{21}^{-1}
\\
&=
\mathcal R_{21}\,
\rho^{(2)}_{\mathrm{op}}(h)\,
(A\otimes\mathbbm 1)\,
\mathcal R_{21}^{-1},
\end{split}
\end{equation}
and likewise
\begin{equation}
\begin{split}
\widetilde A\,\rho^{(2)}(h)
&=
\mathcal R_{21}(A\otimes\mathbbm 1)
\mathcal R_{21}^{-1}\rho^{(2)}(h)
\\
&=
\mathcal R_{21}
(A\otimes\mathbbm 1)
\rho^{(2)}_{\mathrm{op}}(h)
\mathcal R_{21}^{-1}.
\end{split}
\end{equation}
Therefore
\begin{equation}
\left[
\rho^{(2)}(h),\widetilde A
\right]
=
\mathcal R_{21}
\left[
\rho^{(2)}_{\mathrm{op}}(h),
A\otimes\mathbbm 1
\right]
\mathcal R_{21}^{-1}
\end{equation}
for \(h=J_z,J_+,J_-\). Thus the commutator with the global coproduct action is equivalent, through the \(R\)-matrix, to the opposite-coproduct action on the
undressed local operator.\\

Let us now formulate the same statement in intrinsic Hopf-algebraic terms. The ordinary commutator identity above is a useful representation-level consequence of quasitriangularity, but the natural notion of covariance for a Hopf-algebra action is the Hopf adjoint action, which involves the coproduct and the antipode. Let \(h\in U_q(\mathfrak{su}(2))\). For any representation \(\pi\) of \(U_q(\mathfrak{su}(2))\) on a vector space \(W\), the induced adjoint action on
\(\mathrm{End}(W)\) is \cite{Kassel1995}
\begin{equation}
h\triangleright_{\pi} T
=
\pi(h_{(1)})\,T\,\pi(S h_{(2)}),
\qquad
T\in \mathrm{End}(W),
\label{eq:HopfAdjointGeneral}
\end{equation}
where \(S\) is the antipode. In the present case we have two representations of \(H\) on the same
two-qubit vector space,
\begin{equation}
\rho^{(2)}(h)=(\rho\otimes\rho)\Delta(h),
\qquad
\rho^{(2)}_{\mathrm{op}}(h)
=
(\rho\otimes\rho)\Delta^{\mathrm{op}}(h).
\end{equation}
We denote the corresponding adjoint actions by
\begin{equation}
h\triangleright_{\Delta} T
=
\rho^{(2)}(h_{(1)})\,T\,\rho^{(2)}(S h_{(2)}),
\label{eq:AdjDelta}
\end{equation}
and
\begin{equation}
h\triangleright_{\Delta^{\mathrm{op}}} T
=
\rho^{(2)}_{\mathrm{op}}(h_{(1)})\,
T\,
\rho^{(2)}_{\mathrm{op}}(S h_{(2)}).
\label{eq:AdjDeltaOp}
\end{equation}
The quasitriangularity relation implies
\begin{equation}
\mathcal R_{21}\,\rho^{(2)}_{\mathrm{op}}(x)
=
\rho^{(2)}(x)\,\mathcal R_{21},
\qquad
x\in U_q(\mathfrak{su}(2)),
\label{eq:R21Intertwiner}
\end{equation}
and therefore also
\begin{equation}
\mathcal R_{21}^{-1}\,\rho^{(2)}(x)
=
\rho^{(2)}_{\mathrm{op}}(x)\,\mathcal R_{21}^{-1}.
\label{eq:R21IntertwinerInverse}
\end{equation}
Applying these identities to the \(R\)-dressed embedding
\[
\widetilde A
=
\mathcal R_{21}(A\otimes\mathbbm 1)\mathcal R_{21}^{-1},
\]
we obtain
\begin{equation}
\begin{split}
h\triangleright_{\Delta}\widetilde A
&=
\rho^{(2)}(h_{(1)})
\,
\mathcal R_{21}(A\otimes\mathbbm 1)\mathcal R_{21}^{-1}
\,
\rho^{(2)}(S h_{(2)})
\\[2mm]
&=
\mathcal R_{21}\,
\rho^{(2)}_{\mathrm{op}}(h_{(1)})
\,
(A\otimes\mathbbm 1)
\,
\rho^{(2)}_{\mathrm{op}}(S h_{(2)})
\,
\mathcal R_{21}^{-1}
\\[2mm]
&=
\mathcal R_{21}
\left[
h\triangleright_{\Delta^{\mathrm{op}}}
(A\otimes\mathbbm 1)
\right]
\mathcal R_{21}^{-1}.
\end{split}
\label{eq:DressedHopfAdjointCovariance}
\end{equation}

Equation \eqref{eq:DressedHopfAdjointCovariance} means that the dressing map is an intertwiner between two adjoint actions on $\mathrm{End}(\mathbb C^2\otimes\mathbb C^2)$. Acting with the Hopf
adjoint action induced by \(\rho^{(2)}\) after dressing gives the same result as first acting with the Hopf adjoint action induced by \(\rho^{(2)}_{\mathrm{op}}\) and then dressing. In this precise sense, the \(R_{21}\)-dressed embedding is covariant under the deformed symmetry.\\

We ask now the natural question: what is the effect of the deformation on local spin measurements? The answer crucially depends on {\it which} local observable we are considering. As we have already noticed, since \(\mathcal{R}_{21}\) is not unitary with respect to the standard inner product on
\(\mathbb C^2\otimes\mathbb C^2\), the dressed embedding of Alice's one-qubit operator $\widetilde A
= \mathcal{R}_{21}(A\otimes \mathbbm 1)\mathcal{R}_{21}^{-1}.$ is not, in general, Hermitian with respect to that inner product even when \(A\) is
Hermitian. 

This, however, only shows that {\it the standard tensor-product inner product is not the one naturally associated with the dressed embedding.} We can in fact introduce the braided inner product
\begin{equation}
(\Phi,\Psi)_\mathcal{R}
:=
\langle \mathcal{R}_{21}^{-1}\Phi\,|\,\mathcal{R}_{21}^{-1}\Psi\rangle,
\end{equation}
or equivalently
\begin{equation}
(\Phi,\Psi)_\mathcal{R}
=
\langle \Phi|\eta_{\mathcal{R}}|\Psi\rangle,
\qquad
\eta_{\mathcal{R}}=(\mathcal{R}_{21}^{-1})^\dagger \mathcal{R}_{21}^{-1}.
\end{equation}
In the standard ordered basis $\mathcal{B}$ we have
\begin{equation}
\eta_{\mathcal R}
=
\begin{pmatrix}
q^{-2}&0&0&0\\
0&1+(q-q^{-1})^2&-(q-q^{-1})&0\\
0&-(q-q^{-1})&1&0\\
0&0&0&q^{-2}
\end{pmatrix}.
\end{equation}
Let us now denote the elements of the basis  $\mathcal{B}$ as
\begin{equation}
e_1=|\uparrow\uparrow\rangle,\qquad
e_2=|\uparrow\downarrow\rangle,\qquad
e_3=|\downarrow\uparrow\rangle,\qquad
e_4=|\downarrow\downarrow\rangle\,.
\end{equation}
For two generic states in $\mathbb{C}^2\otimes \mathbb{C}^2$
\begin{equation}
|\Phi\rangle
=
a e_1+b e_2+c e_3+d e_4,
\qquad
|\Psi\rangle
=
a' e_1+b' e_2+c' e_3+d' e_4,
\end{equation}
the braided inner product is explicitly
\begin{equation}
\begin{split}
(\Phi,\Psi)_{\mathcal R}
={}&
q^{-2}a^*a'
+
\left[1+(q-q^{-1})^2\right]b^*b'
\\
&-
(q-q^{-1})b^*c'
-
(q-q^{-1})c^*b'
+
c^*c'
+
q^{-2}d^*d' .
\end{split}
\end{equation}

With respect to this inner product, all dressed images of ordinary one-qubit Hermitian operators are Hermitian. Indeed, if \(A=A^\dagger\) in the standard one-qubit Hilbert space, then
\begin{equation}
\begin{split}
(\Phi,\widetilde A\Psi)_\mathcal{R}
&=
\langle \mathcal{R}_{21}^{-1}\Phi\,|\,(A\otimes\mathbbm 1)\mathcal{R}_{21}^{-1}\Psi\rangle
\\
&=
\langle (A\otimes\mathbbm 1)\mathcal{R}_{21}^{-1}\Phi\,|\,\mathcal{R}_{21}^{-1}\Psi\rangle
\\
&=
(\widetilde A\Phi,\Psi)_\mathcal{R}.
\end{split}
\end{equation}

It should be noticed that the standard ordered basis \(\{e_i\}\) is not orthonormal with respect to \((\cdot,\cdot)_\mathcal{R}\). The corresponding dressed orthonormal basis is given by the map 
\begin{equation}
f_i:=\mathcal{R}_{21} e_i,
\end{equation}
explicitly $f_1=q|\uparrow\uparrow\rangle,\,\, f_2=
|\uparrow\downarrow\rangle
+
(q-q^{-1})|\downarrow\uparrow\rangle,\,\, f_3=
|\downarrow\uparrow\rangle,\,\, f_4=q|\downarrow\downarrow\rangle$. By construction, these states are otrhonormal with respect to th ebraided inner product
\begin{equation}
(f_i,f_j)_\mathcal{R}
=
\langle \mathcal{R}_{21}^{-1}f_i\,|\,\mathcal{R}_{21}^{-1}f_j\rangle
=
\langle e_i|e_j\rangle
=
\delta_{ij}.
\end{equation}
Thus the vectors \(f_i\), rather than the original product vectors \(e_i\), are the analogue of the standard orthonormal ordered basis in the braided Hilbert-space description.

If we now look at the standard Alice projectors
\begin{equation}\label{aliceloc}
P_\pm^{(A)}
=
P_\pm\otimes\mathbbm 1,
\qquad
P_+=|\uparrow\rangle\langle\uparrow|,
\qquad
P_-=|\downarrow\rangle\langle\downarrow|,
\end{equation}
their dressed counterparts are
\begin{equation}
\widetilde P_\pm^{(A)}
=
\mathcal{R}_{21} P_\pm^{(A)} \mathcal{R}_{21}^{-1}\,.
\end{equation}
These are orthogonal projectors with respect to the braided inner product:
\begin{equation}
\left(\widetilde P_\pm^{(A)}\right)^\ddagger
=
\widetilde P_\pm^{(A)}.
\end{equation}
where $\ddagger$ denotes hermitian conjugation with respect to the braided inner product and
\begin{equation}
\left(\widetilde P_\pm^{(A)}\right)^2
=
\widetilde P_\pm^{(A)},
\qquad
\widetilde P_+^{(A)}+\widetilde P_-^{(A)}=\mathbbm 1.
\end{equation}

We can therefore define a corresponding Born rule and post-measurement state using \((\cdot,\cdot)_\mathcal{R}\)
\begin{equation}\label{braidborn}
p_\pm^{(\mathcal{R})}(\Psi)
=
\frac{
(\Psi,\widetilde P_\pm^{(A)}\Psi)_\mathcal{R}
}{
(\Psi,\Psi)_\mathcal{R}
},
\end{equation}
and
\begin{equation}\label{braidstateup}
|\Psi_\pm\rangle
=
\frac{\widetilde P_\pm^{(A)}|\Psi\rangle
}{\sqrt{(\Psi,\widetilde P_\pm^{(A)}\Psi)_\mathcal{R}}}.
\end{equation}

We observe, at this point, that we are naturally led to consider two different notions of measurement: the one associated with the braided inner product in which probabilities of outcomes for measurements of braided quantum group covariant local observables \eqref{eq:dresseddef} are given by the braided Born rule \eqref{braidborn} (and associated state update \eqref{braidstateup}), and the one associated to the standard inner product on \(\mathbb C^2\otimes\mathbb C^2\), for which Alice's spin projectors are given by \eqref{aliceloc}
with corresponding local spin observable 
\begin{equation}
J_z^{(A)}
=
\frac12
\left(
P_+^{(A)}-P_-^{(A)}
\right)
=
J_z\otimes \mathbbm 1.
\end{equation}
and the standard Born rule 
\begin{equation}
p_\pm^{\mathrm{loc}}(\Psi)
=
\frac{
\langle\Psi|P_\pm^{(A)}|\Psi\rangle
}{
\langle\Psi|\Psi\rangle
}.
\end{equation}

We can now compare explicitly the measurement statistics obtained from two different notions of locality. Let us restrict attention to the zero-weight subspace of states with total $J_z$ -eigenvalue zero, spanned by
\(e_2,e_3\), and write (in the following we take \(q>0\))
\begin{equation}
|\Psi\rangle=\alpha e_2+\beta e_3 .
\end{equation}
For a standard local Alice measurement one finds
\begin{equation}
p_+^{\mathrm{loc}}(\Psi)
=
\frac{|\alpha|^2}{|\alpha|^2+|\beta|^2},
\qquad
p_-^{\mathrm{loc}}(\Psi)
=
\frac{|\beta|^2}{|\alpha|^2+|\beta|^2},
\end{equation}
and thus
\begin{equation}
\langle J_z^{(A)}\rangle_{\mathrm{loc},\Psi}
=
\frac12
\frac{
|\alpha|^2-|\beta|^2
}{
|\alpha|^2+|\beta|^2
}.
\end{equation}

For the braided-local Alice measurement, instead, using
\begin{equation}
\mathcal{R}_{21}^{-1}e_2=e_2-(q-q^{-1})e_3,
\qquad
\mathcal{R}_{21}^{-1}e_3=e_3,
\end{equation}
we obtain
\begin{equation}
\mathcal{R}_{21}^{-1}|\Psi\rangle
=
\alpha e_2+
\left(\beta-(q-q^{-1})\alpha\right)e_3,
\end{equation}
from which
\begin{equation}
p_+^{(\mathcal{R})}(\Psi)
=
\frac{
|\alpha|^2
}{
|\alpha|^2+
|\beta-(q-q^{-1})\alpha|^2
},
\end{equation}
and
\begin{equation}
p_-^{(\mathcal{R})}(\Psi)
=
\frac{
|\beta-(q-q^{-1})\alpha|^2
}{
|\alpha|^2+
|\beta-(q-q^{-1})\alpha|^2
}.
\end{equation}
The corresponding braided-local expectation value is
\begin{equation}
\langle \widetilde J_z^{(A)}\rangle_{\mathcal{R},\Psi}
=
\frac12
\frac{
|\alpha|^2-
|\beta-(q-q^{-1})\alpha|^2
}{
|\alpha|^2+
|\beta-(q-q^{-1})\alpha|^2
}.
\end{equation}
This shows that standard-local and braided-local measurements are, in general, operationally distinguishable. They coincide at \(q=1\), but for
\(q\neq1\) they probe different tensor-product structures.\\

As an illustrative example let us consider first the ordinary Bell singlet
\begin{equation}
|\psi_0\rangle
=
\frac{1}{\sqrt2}(e_2-e_3).
\end{equation}
For the standard local Alice measurement,
\begin{equation}
p_+^{\mathrm{loc}}(\psi_0)
=
p_-^{\mathrm{loc}}(\psi_0)
=
\frac12,
\end{equation}
and therefore
\begin{equation}
\langle J_z^{(A)}\rangle_{\mathrm{loc},\psi_0}=0.
\end{equation}
However, if the same vector \(|\psi_0\rangle\) is measured using the braided-local prescription, then
\begin{equation}\label{braidsinglet}
\mathcal{R}_{21}^{-1}|\psi_0\rangle
=
\frac{1}{\sqrt2}
\left[
e_2-\left(1+q-q^{-1}\right)e_3
\right],
\end{equation}
thus
\begin{equation}
p_+^{(\mathcal{R})}(\psi_0)
=
\frac{
1
}{
1+\left(1+q-q^{-1}\right)^2
},
\end{equation}
and
\begin{equation}
p_-^{(\mathcal{R})}(\psi_0)
=
\frac{
\left(1+q-q^{-1}\right)^2
}{
1+\left(1+q-q^{-1}\right)^2
}.
\end{equation}
The braided-local expectation value is therefore
\begin{equation}
\langle \widetilde J_z^{(A)}\rangle_{\mathcal{R},\psi_0}
=
\frac12
\frac{
1-\left(1+q-q^{-1}\right)^2
}{
1+\left(1+q-q^{-1}\right)^2
}.
\end{equation}
This is generically nonzero for \(q\neq1\), and in particular is nonzero for small deformations around the undeformed point. 
There is an important consistency check. If one transforms not only the observable, but also the state, then no physical statistics change. Namely, defining
\[
|\psi_{0}\rangle_\mathcal{R}:=\mathcal R_{21}|\psi_0\rangle,
\]
one has
\[
p_\pm^{(\mathcal R)}(\psi_{0,\mathcal R})
=
p_\pm^{\mathrm{loc}}(\psi_0)
=
\frac12.
\]
Thus, transforming both states and observables is merely a change of Hilbert-space description. Operational differences arise when one compares different measurement prescriptions on the same state.\\

Let us now consider the invariant \(q\)-singlet state \eqref{qsingl}
\begin{equation}
|\psi_q\rangle
=
\frac{1}{\sqrt{1+q^{-2}}}
\left(
e_2-q^{-1}e_3
\right).
\end{equation}
For the standard local Alice measurement,
\begin{equation}
p_+^{\mathrm{loc}}(\psi_q)
=
\frac{1}{1+q^{-2}}
=
\frac{q^2}{1+q^2},
\end{equation}
and
\begin{equation}
p_-^{\mathrm{loc}}(\psi_q)
=
\frac{q^{-2}}{1+q^{-2}}
=
\frac{1}{1+q^2}.
\end{equation}
Therefore
\begin{equation}
\langle J_z^{(A)}\rangle_{\mathrm{loc},\psi_q}
=
\frac12
\frac{q^2-1}{q^2+1}.
\end{equation}
Thus a standard tensor-product detector sees a local \(z\)-bias in the
coproduct-invariant singlet whenever \(q\neq1\).

For the braided-local measurement we use
\begin{equation}
\mathcal{R}_{21}^{-1}|\psi_q\rangle
=
\frac{1}{\sqrt{1+q^{-2}}}
\left(
e_2-qe_3
\right),
\end{equation}
from which we obtain
\begin{equation}
p_+^{(\mathcal{R})}(\psi_q)
=
\frac{1}{1+q^2},
\end{equation}
and
\begin{equation}
p_-^{(\mathcal{R})}(\psi_q)
=
\frac{q^2}{1+q^2}.
\end{equation}
The corresponding braided-local expectation value is
\begin{equation}
\langle \widetilde J_z^{(A)}\rangle_{\mathcal{R},\psi_q}
=
\frac12
\frac{1-q^2}{1+q^2}.
\end{equation}
Therefore, for the \(q\)-singlet, we have
\begin{equation}
\langle \widetilde J_z^{(A)}\rangle_{\mathcal{R},\psi_q}
=
-
\langle J_z^{(A)}\rangle_{\mathrm{loc},\psi_q}.
\end{equation}

The comparison can be summarized as follows:\\

\begin{equation}
\begin{array}{c|c|c|c|c}
\text{state}
&
p_+^{\mathrm{loc}}
&
p_-^{\mathrm{loc}}
&
p_+^{(\mathcal R)}
&
p_-^{(\mathcal R)}
\\[2mm]
\hline
\displaystyle
|\psi_0\rangle=\frac{1}{\sqrt2}(e_2-e_3)
&
\displaystyle
\frac12
&
\displaystyle
\frac12
&
\displaystyle
\frac{1}{1+\left(1+q-q^{-1}\right)^2}
&
\displaystyle
\frac{\left(1+q-q^{-1}\right)^2}
{1+\left(1+q-q^{-1}\right)^2}
\\[5mm]
\displaystyle
|\psi_q\rangle
=
\frac{1}{\sqrt{1+q^{-2}}}(e_2-q^{-1}e_3)
&
\displaystyle
\frac{q^2}{1+q^2}
&
\displaystyle
\frac{1}{1+q^2}
&
\displaystyle
\frac{1}{1+q^2}
&
\displaystyle
\frac{q^2}{1+q^2}
\end{array}
\end{equation}
\\

The interpretation is the following. The standard local observable \(J_z\otimes\mathbbm 1\) describes what an ordinary tensor-product qubit
detector coupled to Alice's factor measures. The braided-local observable
\(\widetilde J_z^{(A)}=\mathcal{R}_{21}(J_z\otimes\mathbbm 1)\mathcal{R}_{21}^{-1}\), together with the braided inner product, describes a detector adapted to the braided tensor structure. These are not the same measurement for \(q\neq1\). Therefore their
probability distributions can be operationally distinguished, provided the two measurement procedures are physically implemented as distinct measurement couplings.

In particular, the coproduct-invariant \(q\)-singlet is invariant under the deformed symmetry, but this does not imply that ordinary local measurements are unbiased. Rather, the local bias is precisely a signature
of probing a \(U_q(su(2))\)-invariant state with a standard tensor-product local detector. The braided-local detector probes the transported tensor structure instead, and for the above convention it sees the reciprocal, oppositely oriented bias.\\

Let us conclude by summarizing the main results of the analysis. We studied local spin measurements on bipartite singlet states when rotational symmetry is described not by an ordinary Lie algebra, but by the
quasitriangular Hopf algebra \(U_q(\mathfrak{su}(2))\). Although the fundamental spin-\(\frac12\) representation of the generators is undeformed as an action on one-particle states, the Hopf-algebraic structures that control covariance, composition and tensor-product
representations are deformed. In particular, the non-trivial coproduct changes the notion of total angular momentum and selects a \(q\)-dependent invariant singlet.

The main physical point is that, in the deformed theory, tha action of symmetry generators on composite systems and local measurements are no longer tied together in the same way as in ordinary quantum mechanics. The singlet state selected by the deformed coproduct is invariant under the quantum-group action, but an ordinary tensor-product detector, coupled to Alice's factor through the standard local observable \(J_z\otimes\mathbbm 1\), sees deformation-dependent local
marginals. Thus the invariant singlet can look locally biased to a detector whose notion of locality is the undeformed tensor-factor one.

The quasitriangular structure provides a second, deformed symmetry-adapted notion of local measurement. By dressing one-site observables with the \(R\)-matrix, one obtains braided-local observables which transform covariantly under the Hopf adjoint action. Since the \(R\)-matrix is not unitary with respect to the standard tensor-product inner product, these dressed observables are not generally Hermitian in the standard Hilbert-space structure. They become Hermitian only after passing to the corresponding braided inner product. In this sense, the braided detector is not simply the same detector written in different notation: it belongs to a different Hilbert-space measurement structure.

The comparison between the standard-local and braided-local prescriptions is therefore operationally meaningful only when the same state is probed by
two different measurement couplings. If states, observables and inner product are all transformed no physical statistics change; this is merely a change of description. By contrast, if the coproduct-invariant singlet is kept fixed and one compares the ordinary tensor-factor detector with the braided-local detector, the two prescriptions lead to different marginal statistics. In the convention used here, the braided-local measurement sees the reciprocal, oppositely oriented bias relative to the standard-local
one. Hence the braided prescription does not restore the usual locally unbiased singlet statistics on the \(q\)-singlet; rather, it reveals that different notions of locality probe different tensor structures.

This clarifies the physical interpretation of the effect. A macroscopic Stern-Gerlach apparatus whose calibration is based on an undeformed classical reference frame naturally implements the standard tensor-factor measurement. A detector adapted to the quantum-group tensor structure would instead implement the braided-local measurement rule. The difference
between the two can be interpreted as a calibration mismatch between an ordinary classical notion of locality and the microscopic tensor structure
selected by the deformed symmetry. This suggests that braided-local observables can be viewed as measurements on bipartite systems performed with apparata whose calibration is adapted to reference frames transforming covariantly under deformed infinitesimal rotations, thereby providing a further realization of the idea of quantum reference frames \cite{Poulin:2006ryq,Giacomini:2017zju,Amelino-Camelia:2022dsj}.

More broadly, the analysis suggests that in Hopf-symmetric quantum systems the operational primitives of quantum information theory should be reconsidered. Standard LOCC protocols, separability criteria and local
operations are defined relative to a chosen tensor-product structure and a chosen class of local subalgebras \cite{AndreadakisZanardi2025TPSGeometry, BartlettRudolphSpekkens2007, KitaevMayersPreskill2004}. In a non-cocommutative Hopf-algebraic setting these choices need not be symmetry-adapted. It is therefore natural to ask whether one should formulate a braided analogue of local operations, classical communication and free transformations, in which local Kraus operators and measurements are dressed by the same relocalization map that defines braided-local observables (see also \cite{XuZhou2023TopologicalCorrelation}).

Several directions remain open. One should identify concrete dynamical models of measurement couplings that realize either the standard-local or the braided-local prescription. One should also clarify how quantum-group
covariant reference frames are related to ordinary classical apparatuses, and whether the bias found here can be understood as an experimentally accessible signature of such a mismatch. Finally, it would be interesting to connect the present measurement-level effect with operator-level nonlocality and entangling power generated by coproduct-defined dynamics.
These questions point toward a broader operational framework for quantum information in systems with deformed, Hopf-algebraic symmetries. We leave these directions for future work.\\

\section*{Acknowledgments}
M.A. acknowledges support from the INFN Iniziativa Specifica QUAGRAP. M.A., G.C. and J.KG. acknowledge the support from the European COST Actions BridgeQG CA23130 and CaLISTA CA21109. M.A. also acknowledges support from the Institute of Theoretical Physics at the University of Wrocław where part of this work was carried out.

\bibliographystyle{uiuchept}
\bibliography{biblio}

@book{biedenharn1995quantum,
  title={Quantum Group Symmetry and Q-tensor Algebras},
  author={Biedenharn, L.C. and Lohe, M.A.},
  isbn={9789810223311},
  lccn={96106412},
  series={G - Reference, Information and Interdisciplinary Subjects Series},
  url={https://books.google.it/books?id=DTlqDQAAQBAJ},
  year={1995},
  publisher={World Scientific}
}

@article{Arzano:2025zsp,
    author = "Arzano, Michele and Chirco, Goffredo",
    title = "{Operator Entanglement from Non-Commutative Symmetries}",
    eprint = "2512.24806",
    journal = "",
    archivePrefix = "arXiv",
    primaryClass = "quant-ph",
    year = "2025"
}

@article{Arzano:2026gng,
    author = "Arzano, Michele and Del Prete, Antonio and Frattulillo, Domenico",
    title = "{Quantum Evolution of Hopf Algebra Hamiltonians}",
    eprint = "2602.07887",
    journal = "",
    archivePrefix = "arXiv",
    primaryClass = "quant-ph",
    year = "2026"
}

@article{Bianchi:2011uq,
    author = "Bianchi, Eugenio and Rovelli, Carlo",
    title = "{A Note on the geometrical interpretation of quantum groups and non-commutative spaces in gravity}",
    eprint = "1105.1898",
    archivePrefix = "arXiv",
    primaryClass = "gr-qc",
    doi = "10.1103/PhysRevD.84.027502",
    journal = "Phys. Rev. D",
    volume = "84",
    pages = "027502",
    year = "2011"
}

@article{Amelino-Camelia:2022dsj,
    author = "Amelino-Camelia, Giovanni and D'Esposito, Vittorio and Fabiano, Giuseppe and Frattulillo, Domenico and Hoehn, Philipp A. and Mercati, Flavio",
    title = "{Quantum Euler angles and agency-dependent space-time}",
    eprint = "2211.11347",
    archivePrefix = "arXiv",
    primaryClass = "gr-qc",
    doi = "10.1093/ptep/ptae015",
    journal = "PTEP",
    volume = "2024",
    number = "3",
    pages = "033A01",
    year = "2024"
}

@article{Addazi:2021xuf,
    author = "Addazi, A. and others",
    title = "{Quantum gravity phenomenology at the dawn of the multi-messenger era{\textemdash}A review}",
    eprint = "2111.05659",
    archivePrefix = "arXiv",
    primaryClass = "hep-ph",
    doi = "10.1016/j.ppnp.2022.103948",
    journal = "Prog. Part. Nucl. Phys.",
    volume = "125",
    pages = "103948",
    year = "2022"
}

@book{Arzano:2021scz,
    author = "Arzano, Michele and Kowalski-Glikman, Jerzy",
    title = "{Deformations of Spacetime Symmetries}: {Gravity, Group-Valued Momenta, and Non-Commutative Fields}",
    doi = "10.1007/978-3-662-63097-6",
    isbn = "978-3-662-63095-2, 978-3-662-63097-6",
    series = "Lecture Notes in Physics",
    volume = "986",
    month = "6",
    year = "2021"
}

@article{Tjin:1991me,
    author = "Tjin, T.",
    title = "{An Introduction to quantized Lie groups and algebras}",
    eprint = "hep-th/9111043",
    archivePrefix = "arXiv",
    reportNumber = "PRINT-91-0499 (AMSTERDAM)",
    doi = "10.1142/S0217751X92002805",
    journal = "Int. J. Mod. Phys. A",
    volume = "7",
    pages = "6175--6213",
    year = "1992"
}

@book{Majid:1996kd,
    author = "Majid, S.",
    title = "{Foundations of quantum group theory}",
    isbn = "978-0-511-83453-0, 978-0-521-64868-4",
    publisher = "Cambridge University Press",
    year = "2011"
}

@article{Meusburger:2003hc,
    author = "Meusburger, C. and Schroers, B. J",
    title = "{The quantisation of Poisson structures arising inChern-Simons theory with gauge group $G \ltimes \mathfrak{g}^*$}",
    eprint = "hep-th/0310218",
    archivePrefix = "arXiv",
    reportNumber = "HWM-03-20, EMPG-03-19",
    doi = "10.4310/ATMP.2003.v7.n6.a3",
    journal = "Adv. Theor. Math. Phys.",
    volume = "7",
    number = "6",
    pages = "1003--1043",
    year = "2003"
}

@article{Cianfrani:2016ogm,
    author = "Cianfrani, Francesco and Kowalski-Glikman, Jerzy and Pranzetti, Daniele and Rosati, Giacomo",
    title = "{Symmetries of quantum spacetime in three dimensions}",
    eprint = "1606.03085",
    archivePrefix = "arXiv",
    primaryClass = "hep-th",
    doi = "10.1103/PhysRevD.94.084044",
    journal = "Phys. Rev. D",
    volume = "94",
    number = "8",
    pages = "084044",
    year = "2016"
}

@article{BartlettRudolphSpekkens2007,
  author  = "Bartlett, Stephen D. and Rudolph, Terry and Spekkens, Robert W.",
  title   = "{Reference frames, superselection rules, and quantum information}",
  journal = "Reviews of Modern Physics",
  volume  = "79",
  number  = "2",
  pages   = "555--609",
  year    = "2007",
  doi     = "10.1103/RevModPhys.79.555",
  url     = "https://link.aps.org/doi/10.1103/RevModPhys.79.555"
}

@article{KitaevMayersPreskill2004,
  author  = "Kitaev, Alexei and Mayers, Dominic and Preskill, John",
  title   = "{Superselection rules and quantum protocols}",
  journal = "Physical Review A",
  volume  = "69",
  number  = "5",
  pages   = "052326",
  year    = "2004",
  doi     = "10.1103/PhysRevA.69.052326",
  url     = "https://link.aps.org/doi/10.1103/PhysRevA.69.052326"
}

@article{AndreadakisZanardi2025TPSGeometry,
  author  = "Andreadakis, Faidon and Zanardi, Paolo",
  title   = "{Tensor Product Structure Geometry under Unitary Channels}",
  journal = "Quantum",
  volume  = "9",
  pages   = "1668",
  year    = "2025",
  doi     = "10.22331/q-2025-03-25-1668",
  eprint  = "2410.02911",
  archivePrefix = "arXiv",
  primaryClass  = "quant-ph",
  url     = "https://quantum-journal.org/papers/q-2025-03-25-1668/"
}

@article{XuZhou2023TopologicalCorrelation,
  author  = "Xu, Cheng-Qian and Zhou, D. L.",
  title   = "{Topological correlation: anyonic states cannot be determined by local operations and classical communication}",
  journal = "Physical Review A",
  volume  = "108",
  number  = "5",
  pages   = "052221",
  year    = "2023",
  doi     = "10.1103/PhysRevA.108.052221",
  eprint  = "2306.03596",
  archivePrefix = "arXiv",
  primaryClass  = "quant-ph",
  url     = "https://arxiv.org/abs/2306.03596"
}

@article{Bais:1998yn,
    author = "Bais, F. A. and Muller, N. M.",
    title = "{Topological field theory and the quantum double of SU(2)}",
    eprint = "hep-th/9804130",
    archivePrefix = "arXiv",
    reportNumber = "UVA-WINS-ITFA-98-07",
    doi = "10.1016/S0550-3213(98)00572-0",
    journal = "Nucl. Phys. B",
    volume = "530",
    pages = "349--400",
    year = "1998"
}

@article{Bais:2002ye,
    author = "Bais, F. A. and Muller, N. M. and Schroers, B. J.",
    title = "{Quantum group symmetry and particle scattering in (2+1)-dimensional quantum gravity}",
    eprint = "hep-th/0205021",
    archivePrefix = "arXiv",
    reportNumber = "HWM-01-45, EMPG-02-07, ITFA-2002-12",
    doi = "10.1016/S0550-3213(02)00572-2",
    journal = "Nucl. Phys. B",
    volume = "640",
    pages = "3--45",
    year = "2002"
}

@article{Amelino-Camelia:2003ezw,
    author = "Amelino-Camelia, Giovanni and Smolin, Lee and Starodubtsev, Artem",
    title = "{Quantum symmetry, the cosmological constant and Planck scale phenomenology}",
    eprint = "hep-th/0306134",
    archivePrefix = "arXiv",
    doi = "10.1088/0264-9381/21/13/002",
    journal = "Class. Quant. Grav.",
    volume = "21",
    pages = "3095--3110",
    year = "2004"
}

@article{Kowalski-Glikman:2008fix,
    author = "Kowalski-Glikman, Jerzy and Starodubtsev, Artem",
    title = "{Effective particle kinematics from Quantum Gravity}",
    eprint = "0808.2613",
    archivePrefix = "arXiv",
    primaryClass = "gr-qc",
    doi = "10.1103/PhysRevD.78.084039",
    journal = "Phys. Rev. D",
    volume = "78",
    pages = "084039",
    year = "2008"
}

@article{Meusburger:2003ta,
    author = "Meusburger, C. and Schroers, B. J.",
    title = "{Poisson structure and symmetry in the Chern-Simons formulation of (2+1)-dimensional gravity}",
    eprint = "gr-qc/0301108",
    archivePrefix = "arXiv",
    reportNumber = "HWM-03-2, EMPG-03-02",
    doi = "10.1088/0264-9381/20/11/318",
    journal = "Class. Quant. Grav.",
    volume = "20",
    pages = "2193--2234",
    year = "2003"
}

@book{ChariPressley1994,
  title     = {A Guide to Quantum Groups},
  author    = {Chari, Vyjayanthi and Pressley, Andrew},
  publisher = {Cambridge University Press},
  year      = {1994}
}

@article{Fiore:2000qj,
    author = "Fiore, Gaetano and Steinacker, Harold and Wess, Julius",
    title = "{Unbraiding the braided tensor product}",
    eprint = "math/0007174",
    archivePrefix = "arXiv",
    reportNumber = "LMU-TPW-00-21, MPI-PHT-2000-28",
    doi = "10.1063/1.1522818",
    journal = "J. Math. Phys.",
    volume = "44",
    pages = "1297--1321",
    year = "2003"
}

@book{KlimykSchmudgen1997,
  author    = {Klimyk, Anatoli and Schm{\"u}dgen, Konrad},
  title     = {Quantum Groups and Their Representations},
  publisher = {Springer},
  year      = {1997},
  doi       = {10.1007/978-3-642-60896-4}
}

@book{Kassel1995,
  author    = {Kassel, Christian},
  title     = {Quantum Groups},
  series    = {Graduate Texts in Mathematics},
  volume    = {155},
  publisher = {Springer},
  year      = {1995},
  doi       = {10.1007/978-1-4612-0783-2}
}

@article{Einstein:1935rr,
  author = {Einstein, A. and Podolsky, B. and Rosen, N.},
  title = {Can Quantum-Mechanical Description of Physical Reality Be Considered Complete?},
  journal = {Phys. Rev.},
  volume = {47},
  pages = {777--780},
  year = {1935},
  doi = {10.1103/PhysRev.47.777}
}

@article{Bell:1964kc,
  author = {Bell, J. S.},
  title = {On the Einstein Podolsky Rosen Paradox},
  journal = {Physics Physique Fizika},
  volume = {1},
  pages = {195--200},
  year = {1964},
  doi = {10.1103/PhysicsPhysiqueFizika.1.195}
}

@article{Horodecki:2009zz,
  author = {Horodecki, Ryszard and Horodecki, Pawel and Horodecki, Michal and Horodecki, Karol},
  title = {Quantum entanglement},
  journal = {Rev. Mod. Phys.},
  volume = {81},
  pages = {865--942},
  year = {2009},
  doi = {10.1103/RevModPhys.81.865}
}

@article{Zanardi:2003zz,
  author = {Zanardi, Paolo and Lidar, Daniel A. and Lloyd, Seth},
  title = {Quantum Tensor Product Structures are Observable Induced},
  journal = {Phys. Rev. Lett.},
  volume = {92},
  pages = {060402},
  year = {2004},
  doi = {10.1103/PhysRevLett.92.060402}
}

@article{Barnum:2004zz,
  author = {Barnum, Howard and Knill, Emanuel and Ortiz, Gerardo and Somma, Rolando and Viola, Lorenza},
  title = {A Subsystem-Independent Generalization of Entanglement},
  journal = {Phys. Rev. Lett.},
  volume = {92},
  pages = {107902},
  year = {2004},
  doi = {10.1103/PhysRevLett.92.107902}
}

@article{Borowiec:2009vb,
    author = "Borowiec, A. and Pachol, A.",
    title = "{Classical basis for kappa-Poincare algebra and doubly special relativity theories}",
    eprint = "0903.5251",
    archivePrefix = "arXiv",
    primaryClass = "hep-th",
    doi = "10.1088/1751-8113/43/4/045203",
    journal = "J. Phys. A",
    volume = "43",
    pages = "045203",
    year = "2010"
}

@article{Giacomini:2017zju,
    author = "Giacomini, Flaminia and Castro-Ruiz, Esteban and Brukner, {\v{C}}aslav",
    title = "{Quantum mechanics and the covariance of physical laws in quantum reference frames}",
    eprint = "1712.07207",
    archivePrefix = "arXiv",
    primaryClass = "quant-ph",
    doi = "10.1038/s41467-018-08155-0",
    journal = "Nature Commun.",
    volume = "10",
    number = "1",
    pages = "494",
    year = "2019"
}

@article{Poulin:2006ryq,
    author = "Poulin, David and Yard, Jon",
    title = "{Dynamics of a quantum reference frame}",
    eprint = "quant-ph/0612126",
    archivePrefix = "arXiv",
    doi = "10.1088/1367-2630/9/5/156",
    journal = "New J. Phys.",
    volume = "9",
    number = "5",
    pages = "156--156",
    year = "2007"
}

\end{document}